\documentclass[conference]{IEEEtran}
\IEEEoverridecommandlockouts
\usepackage{cite}
\usepackage{amsmath,amssymb,amsfonts}
\usepackage{algorithmic}
\usepackage{graphicx}
\usepackage{textcomp}
\usepackage{xcolor}
\usepackage{caption}
\usepackage{subcaption}
\def\BibTeX{{\rm B\kern-.05em{\sc i\kern-.025em b}\kern-.08em
    T\kern-.1667em\lower.7ex\hbox{E}\kern-.125emX}}

\begin{document}

\title{Synchronizing Full-Body Avatar Transforms with WebRTC DataChannel on Educational Metaverse\\}

\author{\IEEEauthorblockN{Yong-Hao Hu}
\IEEEauthorblockA{\textit{Virtual Reality Educational Research Center} \\
\textit{The University of Tokyo}\\
Tokyo, Japan \\
yh-haoareyou@cyber.t.u-tokyo.ac.jp}
\and
\IEEEauthorblockN{Kenichiro Ito}
\IEEEauthorblockA{\textit{Information Technology Center} \\
\textit{The University of Tokyo}\\
Tokyo, Japan \\
ito@ecc.u-tokyo.ac.jp}
\and
\IEEEauthorblockN{Ayumi Igarashi}
\IEEEauthorblockA{\textit{Graduate School of Medicine} \\
\textit{The University of Tokyo}\\
Tokyo, Japan \\
aigarashi-tky@g.ecc.u-tokyo.ac.jp}
}

\maketitle

\begin{abstract}
Full-body avatars are suggested to be beneficial for communication in virtual environments, and consistency between users' voices and gestures is considered essential to ensure communication quality.
This paper propose extending the functionality of a web-based VR platform to support the use of full-body avatars and delegated avatar transforms synchronization to WebRTC DataChannel to enhance the consistency between voices and gestures.
Finally, we conducted a preliminary validation to confirm the consistency.
\end{abstract}


\section{Introduction}

'Metaverse' is commonly defined as 3D virtual environments where interactions among users occur and are accessible via computers or Virtual Reality (VR) devices, and it has found utility across diverse areas, including education.
We have started developing an educational platform in metaverse based on Mozilla Hubs, an open-source web-based VR platform \cite{mypaper}.

\begin{figure}[t]
    \begin{subfigure}[t]{.45\columnwidth}
      \centering
      \includegraphics[width=0.65\linewidth]{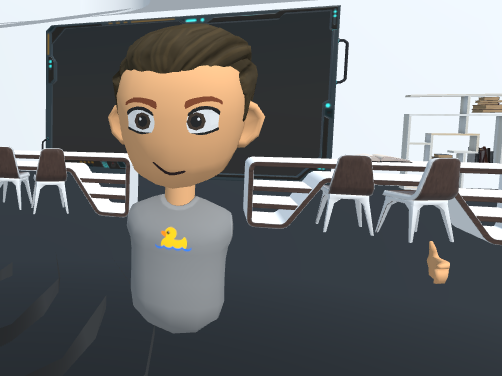}
      \caption{Simplified avatar supported by original implementation}
      \label{half-body}
    \end{subfigure}
    \hfill
    \begin{subfigure}[t]{.45\columnwidth}
      \centering
      \includegraphics[width=0.65\linewidth]{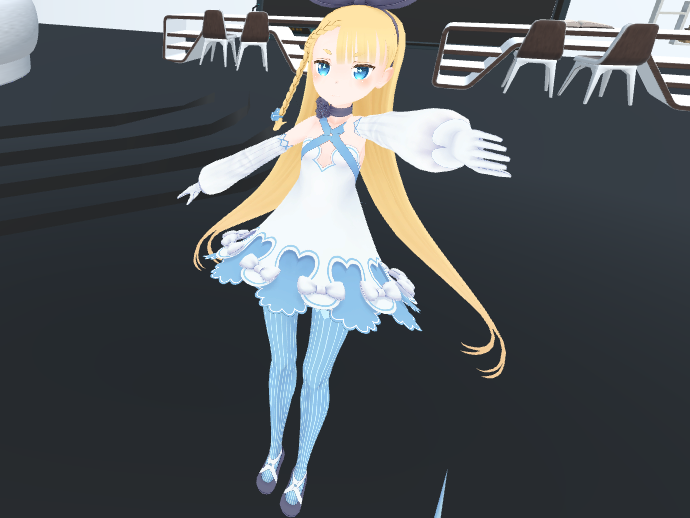}
      \caption{Full-body avatar used in our proposed implementation}
      \label{full-body}
    \end{subfigure}
    \caption{Screenshots by the Observer during validation.}
    \label{avatars}
\end{figure}

'Avatar' refers to a character that represents and is controlled by a user in a virtual environment.
Currently, Mozilla Hubs only supports avatars having a simplified upper body (Fig. \ref{half-body}), which may reduce computation costs on the client side, ensuring Mozilla Hubs' high accessibility even on low-end devices.
On the other hand, it limits the conveyance of non-verbal information through body gestures.
We believed that it is also essential for effective communication, and the use of full-body avatars that possess complete body parts like a real human (Fig. \ref{full-body}) can compensate this function.

In fact, previous studies have shown how full-body avatars benefit communication in virtual environments \cite{Cassell1999FullyEC, Wang2019EffectOF}.
Similarly, full-body avatars can improve sense of presence \cite{Heidicker2017InfluenceOA}, which in turn play an important role in learning \cite{presense_in_training}.
Following these findings, we decided to integrate full-body avatars to the Mozilla-Hubs-based platform we are developing.

In addition, Mozilla Hubs currently synchronizes avatar transforms (changes in the position, orientation, or size of an avatar's bones) through WebSocket on a mesh network, while we considered WebRTC DataChannel more suitable to synchronize avatar transforms due to security concerns and consistency between users' voices and gestures. 
In this study, we aimed to expand Mozilla Hubs' implementation to enable the use of full-body avatars and to have the full-body avatar transforms synchronized by WebRTC DataChannel.

\section{Proof of Concept}

\subsection{Implementation}



\subsubsection{Accommodate Full-body Avatars in Mozilla Hubs}

In the original implementation in Mozilla Hubs, avatars are hard-coded to contain limited bone names and hierarchy \footnote{https://github.com/MozillaReality/hubs-avatar-pipelines}, and other avatars with different skeletons, including full-body avatars, are usually neither operable nor rendered properly.
To address this, we prepared a bone mapping function that maps the bones by checking the similarity of their name with their corresponding body parts (e.g. LowerArm.R is mapped to right elbow).
Then we implemented Cyclic Coordinate Descent Inverse Kinematics \cite{CCDIK} so that a full-body avatar can still reflect its user's poses naturally with limited inputs.

\subsubsection{Synchronization of Full-body Avatar Transforms by WebRTC DataChannel}

WebRTC (Web Real-Time Communication) \footnote{https://webrtc.org/} enables real-time data transmission between web browsers without passing through a central server.
WebRTC DataChannel is capable of transmitting text or binary data, implemented on top of UDP while remaining reliability similar to TCP, and incorporating DTLS to encode the data transmission.

In our previous study \cite{mypaper}, we introduced an alternative WebRTC SFU solution into Mozilla Hubs to enhance its audio transmission, and we also delegated the transmission of avatar transforms to DataChannel.
In this current study, we continued to employ this delegation and extended it to encompass full-body avatars for the following reasons.

Firstly, latency on the synchronization of voices and gestures may cause inconsistency between verbal and non-verbal cues, impacting the communication quality and effectiveness.
We argued that real-time alignment between verbal and non-verbal cues with minimal latency should be prioritized, and DataChannel holds the potential to address this concern.

Moreover, within the metaverse, avatar transforms can be regarded as sensitive personal data, especially when using full-body avatars which reproduce a user's real body poses, revealing more about the user's identity\cite{291259}.
Leveraging WebRTC DataChannel allows us to encode avatar transforms transmission and avoid routing through central servers, thereby enhancing higher security.

\subsection{Preliminary Validation in Data Transmission Latency}

A preliminary validation was conducted to measure the transmission latency between audio and avatar transforms within both the original Mozilla Hubs architecture and our proposed one.
We hypothesized that our implementation brings lower latency of avatar transforms from audio, resulting in higher consistency between audio and avatar transforms.

\begin{figure}[t]
\centerline{\includegraphics[width=.75\linewidth]{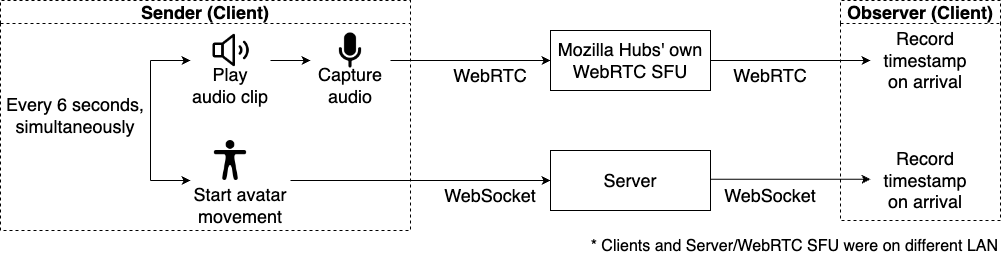}}
\caption{Validation Settings with Original Implementation.}
\label{test-original}
\end{figure}

\begin{figure}[t]
\centerline{\includegraphics[width=.75\linewidth]{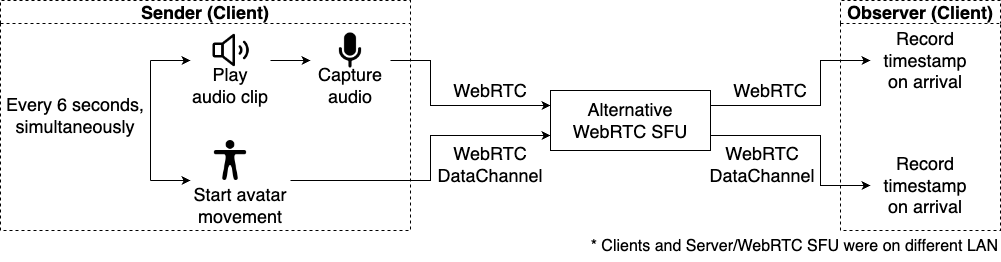}}
\caption{Validation Settings with Proposed Implementation.}
\label{test-proposed}
\end{figure}

One Apple MacBook serving as the Sender and one Windows 10 Desktop Computer serving as the Observer, two devices were connected to the same room on self-hosted Mozilla Hubs (Fig.\ref{test-original} and \ref{test-proposed}).
For 5 minutes, the Sender repeatedly played a 3 second audio clip and synchronized circular movements of its avatar's left hand.
The avatar's left hand positions were transmitted to the Observer, while the audio clip was captured by the Sender's microphone and also transmitted.
Timestamps were recorded whenever the Observer received the audio or observed changes in the Sender's avatar.


\begin{figure}[t]
\centerline{\includegraphics[width=.65\linewidth]{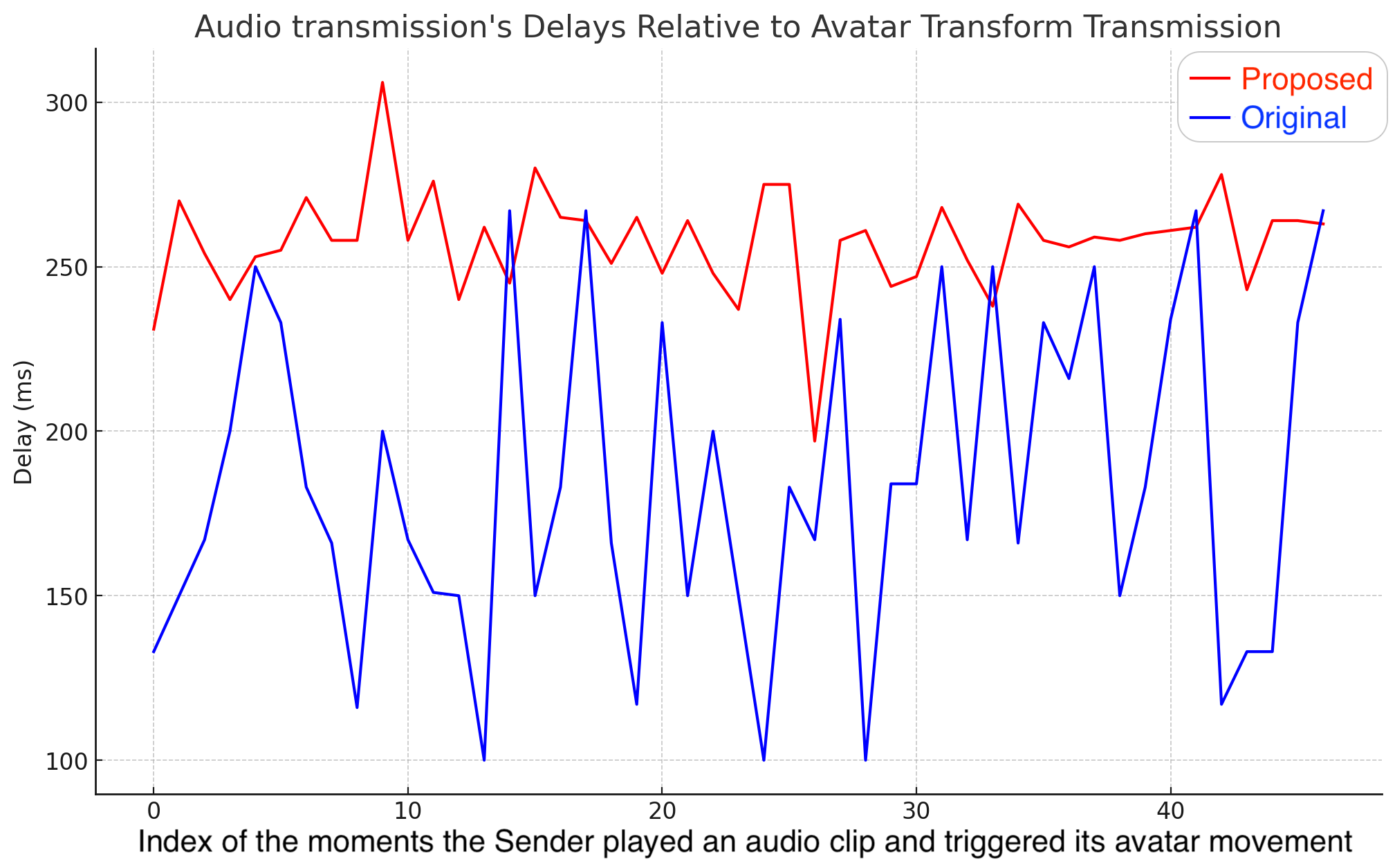}}
\caption{Latency between audio and avatar transforms over time.}
\label{delay}
\end{figure}

Subsequently, the transmission latency was calculated between avatar transforms and audio using the recorded timestamps, and found that avatar transforms were transmitted faster than audio in both conditions (Fig. \ref{delay}).
In our proposed implementation, the average latency was 257.64 ms (SD = 16.10 ms), while in the original implementation, it was 184.04 ms (SD = 49.81 ms).
This suggests that, compared to the original implementation, our approach results in higher dispersion between audio and avatar transforms due to either slower audio transmission or faster synchronization of avatar transforms.
Further investigation is needed to clarify this.

Higher variation was also observed in latency in the original implementation, as evident from the graph and the larger standard deviation, indicating greater stability in our proposed implementation, which can also contribute to higher consistency between audio and avatar transforms.

Regarding the limitations of this validation, it is worth noting that the avatar's movement was triggered when the audio clip was played, not precisely when the Sender started the audio data transmission.
This discrepancy might had contributed to slower audio data arrivals at the Observer side.

Lastly, it is essential to acknowledge that this preliminary validation involved only two devices in the same room.
In rooms accommodating more users, latency in both audio transmission and avatar transforms synchronization may become more obvious, allowing for more meaningful comparisons.

\section{Conclusion}
We extended the functionality of Mozilla Hubs to support the use of full-body avatars and delegated full-body avatar transforms synchronization to WebRTC DataChannel.
The result of a preliminary validation failed to demonstrate a more accurate synchronization but indicated more consistent time differentials between audio and avatar transforms in our implementation.
To gain a clearer understanding of the latency improvement, further investigation is required under higher client load, and we are also planning an usability assessment for our implementation within an educational context.

\section*{Acknowledgment}
This work was partially supported by the following grants: JST Grant Number JPMJPF2202.

\bibliographystyle{IEEEtran}
\bibliography{references}

\begin{thebibliography}{1}
\providecommand{\url}[1]{#1}
\csname url@samestyle\endcsname
\providecommand{\newblock}{\relax}
\providecommand{\bibinfo}[2]{#2}
\providecommand{\BIBentrySTDinterwordspacing}{\spaceskip=0pt\relax}
\providecommand{\BIBentryALTinterwordstretchfactor}{4}
\providecommand{\BIBentryALTinterwordspacing}{\spaceskip=\fontdimen2\font plus
\BIBentryALTinterwordstretchfactor\fontdimen3\font minus \fontdimen4\font\relax}
\providecommand{\BIBforeignlanguage}[2]{{%
\expandafter\ifx\csname l@#1\endcsname\relax
\typeout{** WARNING: IEEEtran.bst: No hyphenation pattern has been}%
\typeout{** loaded for the language `#1'. Using the pattern for}%
\typeout{** the default language instead.}%
\else
\language=\csname l@#1\endcsname
\fi
#2}}
\providecommand{\BIBdecl}{\relax}
\BIBdecl

\bibitem{mypaper}
Y.-H. Hu, K.~Ito, and A.~Igarashi, ``Improving real-time communication for educational metaverse by alternative webrtc sfu and delegating transmission of avatar transform,'' in \emph{Proceedings of International Conference on Consumer Electronics-Taiwan 2023 (ICCETW)}, July 2023.

\bibitem{Cassell1999FullyEC}
J.~Cassell and H.~H. Vilhj{\'a}lmsson, ``Fully embodied conversational avatars: Making communicative behaviors autonomous,'' \emph{Autonomous Agents and Multi-Agent Systems}, vol.~2, pp. 45--64, 1999.

\bibitem{Wang2019EffectOF}
T.-Y. Wang, Y.~Sato, M.~Otsuki, H.~Kuzuoka, and Y.~Suzuki, ``Effect of full body avatar in augmented reality remote collaboration,'' \emph{2019 IEEE Conference on Virtual Reality and 3D User Interfaces (VR)}, pp. 1221--1222, 2019.

\bibitem{Heidicker2017InfluenceOA}
P.~Heidicker, E.~Langbehn, and F.~Steinicke, ``Influence of avatar appearance on presence in social vr,'' \emph{2017 IEEE Symposium on 3D User Interfaces (3DUI)}, pp. 233--234, 2017.

\bibitem{presense_in_training}
F.~Mantovani and G.~Castelnuovo, ``The sense of presence in virtual training: Enhancing skills acquisition and transfer of knowledge through learning experience in virtual environments,'' in \emph{Being there: Concepts, effects and measurements of user presence in synthetic environments.}, G.~Riva, F.~Davide, and W.~A. IJsselsteijn, Eds.\hskip 1em plus 0.5em minus 0.4em\relax IOS Press, 01 2003, ch.~11, pp. 167--181.

\bibitem{CCDIK}
B.~Kenwright, ``Inverse kinematics – cyclic coordinate descent (ccd),'' \emph{Journal of Graphics Tools}, vol.~16, no.~4, pp. 177--217, 2012.

\bibitem{291259}
V.~Nair, W.~Guo, J.~Mattern, R.~Wang, J.~F. O{\textquoteright}Brien, L.~Rosenberg, and D.~Song, ``Unique identification of 50,000+ virtual reality users from head \& hand motion data,'' in \emph{32nd USENIX Security Symposium (USENIX Security 23)}, Anaheim, CA, Aug. 2023, pp. 895--910.

\end{thebibliography}

\end{document}